\documentclass[twocolumn,showpacs,preprintnumbers,amsmath,amssymb,epsfig]{revtex4}

\usepackage{epsfig}
\newcommand{\ccon}{^{\ast}} %dc E\ccon c'est le complexe conjugué de E
\newcommand{\Exp}[1]{\textit{\large{e}}^{\,#1}}
\renewcommand{\Re}{{\mathrm{\,Re\,}}}

\newcommand{\tps}{\tau}
    \newcommand{\apeu}{{a_e}_1}
    \newcommand{\aped}{{a_e}_2}
    \newcommand{\nivf}{\left| f \right\rangle}
    \newcommand{\omeg}{\omega_{eg}}
    \newcommand{\mueg}{\mu_{eg}}
    \newcommand{\fluo}[1]{S^{\,#1}} %\fluo{} pour S, \fluo{\theta} pour S^theta

\newcommand{\be}{\begin{eqnarray}}
\newcommand{\ee}{\end{eqnarray}}

\begin{document}

%\title{Atomic spirograph : measurement of the excited state wave function of a two-level system using coherent transients}
\title{Quantum state measurement using coherent transients}
\author{ Antoine Monmayrant, B\'eatrice Chatel,
and Bertrand Girard}
\affiliation{Laboratoire Collisions,
Agr\'egats, R\'eactivit\'e (UMR 5589 CNRS-UPS), IRSAMC, Universit\'e
Paul Sabatier Toulouse 3, 31062 Toulouse cedex 9, France}
%\address{Laboratoire Collisions, Agr\'egats, R\'eactivit\'e (CNRS UMR 5589), IRSAMC, Univ. P. Sabatier, 31062 Toulouse, France\\}

\date{\today}

\begin{abstract}
We present the principle and experimental demonstration of Time
Resolved Quantum State Holography. The quantum state of an excited
state interacting with an ultrashort chirped laser pulse is measured
during this interaction. This has been obtained by manipulating
coherent transients created by the interaction of femtosecond shaped
pulses and rubidium atoms.

\end{abstract}


\maketitle

%\section{Introduction}
Quantum state measurement is a central issue of fundamental
importance to quantum mechanics \cite{Raymer97,Schleichbook01}.
Since only probabilities can be predicted by quantum mechanics, the
phase of a wave function seems at first sight to carry no
information. However, relative phases between quantum states are
crucial in many circumstances such as prediction of the free or
driven evolution of the system, or the measurements of quantities
(observables) related to the superposition of quantum states with
different energies.

Several examples of quantum phase measurements of states created by
ultrashort pulses are based on interferences between an unknown wave
function and a "reference" wave function. These wave functions are
created by a sequence of two ultrashort pulses (an unknown pulse and
a reference pulse). In quantum state holography, the quantum state
created by the unknown pulse is deduced either by time- and
frequency- integrated fluorescence measured as a function of the
delay \cite{SchleichShapiro98PRL,SchleichShapiro99PRA}, or by
measuring the population of each eigenstate for different values of
the relative phases \cite{Yeazell_holography_97}, or the amplitude
of fluctuations when the delay is randomly fluctuating
\cite{Bucksbaum_QSholography_PRL98,Bucksbaum_holography_Nature99}.
In fluorescence tomography, position probability distributions are
measured as a function of time \cite{Walmsley95,Raymer96}. For
instance, the dispersed fluorescence emitted by an oscillating
nuclear wave packet in a diatomic molecule provides the position
distribution through the Franck-Condon principle \cite{Walmsley95}.
Alternatively the induced dipole can be obtained from heterodyne
measurement \cite{walmsleyPRA99heterodyne}. More recently the
internuclear quantum states of dissociating molecules have been
elegantly measured by tomography using velocity map imaging
\cite{Stapelfeldt03I2tomography}.

In all these examples, the quantum state is first prepared and then
measured in a second step. In the work reported here, the quantum
state is measured {\it during} the interaction with the unknown
laser pulse. Its evolution is thus recorded in real time.

When the matter-light interaction is in the linear regime, the final
state populations can be entirely deduced from the power spectrum.
This is for instance the case for a one-photon transition in the
weak field regime. However, the phase of the wave function is
sensitive to the various phases of the electric field. This can have
important consequences for applications where a subsequent
excitation is performed, in particular when coherent superpositions
are involved.

The transient evolution of excited state population is also strongly
dependent on the details of the pulse shape. In particular,
non-resonant contributions are as important as resonant ones. As an
intuitive illustration of this statement, the transient response to
a non-resonant excitation follows the electric field temporal
envelope, independently of its spectrum. For instance, simply
changing the pulse duration changes this transient response.

\begin{figure}[!ht]
\begin{center}
\epsfig{figure=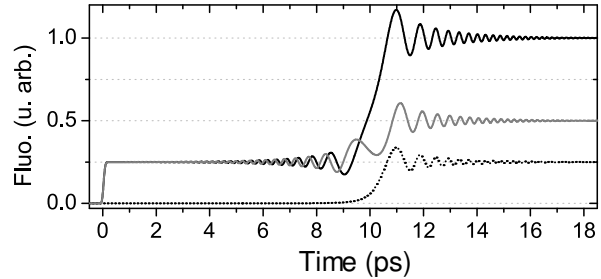,width=0.9 \linewidth} \caption[CT]
{Simulation of Coherent Transients resulting from the excitation of
the atom by a FT limited pulse (at time t=0) followed by a chirped
pulse (centered at t=10ps). For $\theta=0$ (black) and
$\theta=\pi/2$ (gray). Dotted line:  "Simple" CTs obtained for a
single pulse at t=10ps. } \label{FigCTtheo}
\end{center}
\end{figure}

A resonant interaction leads to radically different behavior.
Fourier Transform (FT) limited pulses produce a step-wise excitation
in the weak-field regime, and Rabi oscillations in the intermediate
and strong field regime. Chirped pulses produce a total population
inversion in the strong field with a final state robust with respect
to small variations of laser parameters \cite{Noordamladder92}.
Chirped pulses in the weak field regime lead to Coherent Transients
(CTs) \cite{Vitanov99,zamith01}, a less intuitive behavior which
illustrates the relative importance of the various stages of the
interaction. The laser frequency sweeps linearly with time and
crosses the resonance. Most of the population transfer occurs at
resonance. The small fraction of excited state amplitude transferred
after resonance leads to strong oscillations due to interferences
between the oscillating atomic dipole and the exciting field (see
Fig. \ref{FigCTtheo}, dotted line). Otherwise, interaction before
resonance results in negligible effects. Similarly, in propagation
experiments, interferences between the field radiated by the atom
and the incoming field leads to interferences which can be used as a
partial analysis of the field such as chirp \cite{Rothenberg86} or
dispersion \cite{Delagnes04} measurements.

The shape of these CTs can be radically changed by using a pulse
shaper \cite{Degert02CTshaped,RbShapingAPB04}. This high sensitivity
of Coherent Transients to slight modifications of the laser pulse
provides possibilities to use CT measurements as a means to
characterize the laser pulse itself
\cite{MonmayrantCT-reconstruction-05} or the quantum state which is
generated. In a simple approach, if the general shape of the laser
pulse is known and only few parameters need to be determined, one
can use a simple adjustment of these parameters to fit the
experimental CTs with the predicted ones \cite{RbShapingAPB04}.
However, one would like to establish a general method providing a
direct inversion from the data to the quantum state.
%Using the CT needs to have a dominant quadratic spectral phase. It can be added to the pulse if necessary.
%The CT provide unique features: As a self reference method, it is sensitive to the relative phases with respect to the phase at the atomic resonance frequency.
One limitation of CTs is that only the part of the pulse after
resonance leads to oscillations which can be used to determine the
shape. Another difficulty is that the measured quantity is related
to the excited state probability whereas one aims to measure
probability amplitudes. In this contribution we show how it is
possible to overcome these difficulties. Several CT measurements are
combined, each with a sequence of two pulses. This sequence consists
of a reference pulse which creates an initial population in the
excited state, and an "unknown" pulse whose effect on the quantum
state is to be measured in real time. The first pulse creates an
initial population in the excited state so that the corresponding
dipole beats with the whole of the second (unknown) pulse. A second
measurement is performed after adding a $\pi /2$ phase shift to the
second pulse. This provides in-phase and in-quadrature measurements
from which the real-time evolution ({\it during the laser
interaction}) of the quantum state can be deduced. In common with
quantum state holography \cite{SchleichShapiro98PRL}, the quantum
state created by the unknown pulse is determined by direct
interferometric comparison with a reference quantum state. However,
in our method, the unknown quantum state is measured at any time
during and after its creation. Moreover, the properties of the first
pulse need not be fully characterized in our method.
%{\it The letter is organized as follows. We first recall the
%principle of these coherent transients and how they are
%manipulated. Then the experimental wavefunction reconstruction is
%presented. Finally the possibility of extending this scheme to the
%electric field measurement (or to vibrational levels) will be
%discussed.}

Consider the resonant interaction between an atomic system and a
sequence of two weak nonoverlapping femtosecond laser pulses ${\cal
E}_1 \left( t \right) $ and ${\cal E}_2 \left( t \right) $. First
order time dependent perturbation theory predicts the probability of
finding the atom in the excited state at any time $\tau$ after the
interaction with the first pulse to be
%during (or
%after) the interaction with the second pulse to be
\begin{eqnarray}\label{cc:eq:cttheta-final}
\fluo{\theta}(\tps)&=&\left|\apeu(\infty)+\Exp{i\theta}\aped(\tps)\right|^2\\
\nonumber&=&\left|\apeu(\infty)\right|^2+\left|\aped(\tps)\right|^2+2\Re\left[\Exp{i\theta}\apeu\ccon(\infty)\aped(\tps)\right]
\end{eqnarray}
% \be \label{amplitude}
% a_e \left( t \right) & = & \frac{\mu _{eg}}{\hbar } \left\{ \frac{1}{2}E\left( \omega _{eg}
% \right)e^{i\phi ( \omega _{eg} )} \right. \nonumber\\
% & + & \left. \frac{i}{2\pi} \wp \int_{ - \infty }^{ + \infty }
% d\omega \frac{e^{i\left[ \left( \omega  - \omega _{eg} \right)t + \phi \left( \omega  \right) \right]} }
% {\omega  - \omega _{eg} }E\left( \omega  \right) \right\}\ee
Here $\theta$ is an arbitrary phase which can be added to the
second pulse with respect to the first one. The probability
amplitude produced by the pulse ${\cal E}_k \left( t \right) $ is
\begin{equation}\label{cc:eq:apet-def}
a_{ek}(t) = -\frac{\mueg}{2i\hbar}\int^{t}_{-\infty}{\cal
E}_k(t')\Exp{i\omeg t'} {dt'}
\end{equation}
where $\omeg$ is the transition frequency and $\mueg$ the dipole
moment matrix element. The excited state probability
$\fluo{\theta}(\tps)$ can be measured in a pump-probe scheme
 with an ultrashort probe pulse tuned on a transition towards another
 excited state $\nivf$ \cite{zamith01}. The probability
amplitude produced by the second pulse $a_{e2}(t)$ can be obtained
by resolving the nonlinear equation array resulting from a set of
two experiments performed with different values of $\theta$. For
instance, $\theta_0$ and $\theta_0+\pi/2$ can be chosen. If the
second pulse is much weaker than the first one, the quadratic term
in $a_{e2}(\tps)$ can be neglected to obtain a simple linear
equation array. Alternatively, a third measurement can be performed
with only the second pulse in order to measure $|a_{e2}(\tps)|^2$
directly.

The two pulse sequence with a well defined phase relationship can
be generated at once by a pulse shaper \cite{pulseshaperRSI04}.
For this purpose, the complex spectral mask \be\label{mask}
  H_\theta
  (\omega)=\{1+\exp[i(\theta+\phi'\Delta \omega+\phi''(\Delta \omega)^2/2)]\}/2
\ee is applied to modify the electric field. Here $\Delta \omega =
\omega-\omega_L$, where $\omega_L$ is the laser carrier frequency.
The first pulse is short (FT limited) and the second one is delayed
by $\phi'$ and strongly chirped by $\phi''$ in order to produce the
CTs. The first pulse creates the required initial excited population
to produce oscillations during the whole second pulse as shown in
Fig. \ref{FigCTtheo}.

\begin{figure}[!ht]
\begin{center}
\epsfig{figure=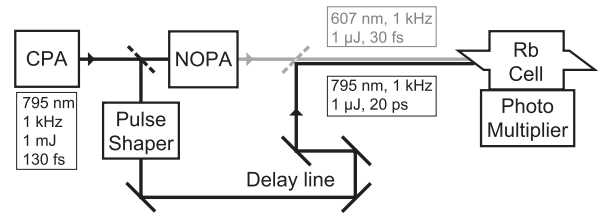,width=0.9\linewidth} \caption[setup]
{Experimental set-up. NOPA : non collinear optical parametric
amplifier; CPA: Chirped Pulse Amplifier} \label{Fig-setup}
\end{center}
\end{figure}

To illustrate this scheme, an experiment has been performed in
atomic rubidium. The experimental set-up is displayed in Fig.
\ref{Fig-setup}.
The 5s-5p(P$_{1/2}$) transition (at 795 nm) is resonantly excited
with a pulse sequence. The transient excited state population is
probed "in real time" on the 5p-(8s, 6d) transitions with an
ultrashort pulse (at 607 nm). The laser system is based on a
conventional Ti:Sapphire laser with chirped pulse amplification
(Spitfire, Spectra Physics) which supplies 1 mJ -130 fs -795 nm
pulses. Half of the beam is used for the pump pulse. The remaining
half seeds a home made Non-collinear Optical Parametric Amplifier
(NOPA) compressed using double pass silica prisms, which delivers
pulses of a few $\rm{\mu J}$, 30 fs (FWHM pulse intensity), centered
around 607 nm. The pump pulse is shaped with a programmable
pulse-shaping device\cite{pulseshaperRSI04}, recombined with the
probe pulse and sent into a sealed rubidium cell.
%All the experiments are performed in the perturbative regime.
The pump-probe signal is detected by monitoring the fluorescence at
420 nm due to the radiative cascade (8s, 6d) - 6p - 5s collected by
a photomultiplier tube as a function of the pump-probe delay. The
pulse shaping device is a 4f set-up composed of one pair each of
reflective gratings and cylindrical mirrors. Its active elements
-two 640 pixels liquid crystal masks- are installed in the common
focal plane of both mirrors. This provides high resolution pulse
shaping in both phase and amplitude \cite{pulseshaperRSI04}. This is
used to generate the shaped pump pulse by applying the function
$H_\theta$ defined above. The laser is centered at resonance
($\omega_L=\omeg$). The delay between the two pulses is equal to
$\phi'= 6 \; {\rm ps}$. The second pulse is strongly chirped
($\phi"=-2.10^5 {\rm \;fs^2}$) to around $10 \; {\rm ps}$ duration.

\begin{figure}[!ht]
\begin{center}
\epsfig{figure=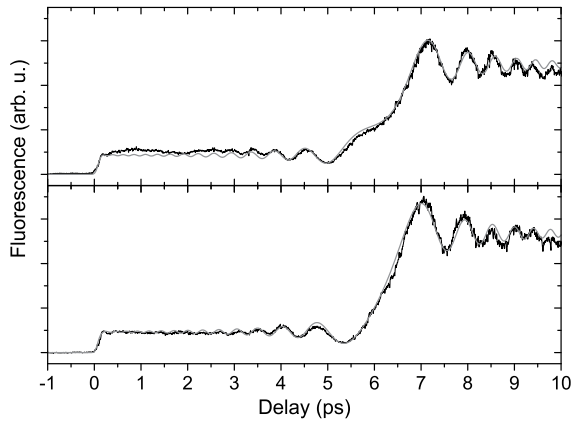,width=1.\linewidth} \caption[CT]
{Coherent Transients resulting from the excitation of the atom by a
FT limited pulse (at time t=0) followed by a chirped pulse (centered
at t=6ps). Lower panel: relative phase between pulses
$\theta=\theta_0 \simeq -0.7 \; {\rm rad}$ and upper panel :
$\theta=\theta_0 + \pi/2$. Solid line : theory; squares :
experiment.} \label{Fig-CT}
\end{center}
\end{figure}
%The agreement is so good that theory and experiment are
%indistinguishable
 Figure \ref{Fig-CT} displays the Coherent Transients obtained with
the two experiments performed with
 $\theta=\theta_0 \simeq -0.7 \; {\rm rad}$ and
$\theta=\theta_0 + \pi/2$. The squares represent the experimental
results which fit perfectly with the resolution of the
Schr\"{o}dinger equation (solid line). The first plateau (for
positive times) is due to the population induced by the first FT
limited pulse. It allows $\left|\apeu(\infty)\right|^2$ to be
determined.

The effects of the second pulse can be divided into three parts. The
strong increase taking place at resonance (around $t=6 \; {\rm ps}$)
is preceded and followed by oscillations resulting from
interferences of off -resonance contributions with the resonant
population. The height of the asymptotic value depends strongly on
the relative phase $\theta$.
\begin{figure}
\begin{center}
\epsfig{figure=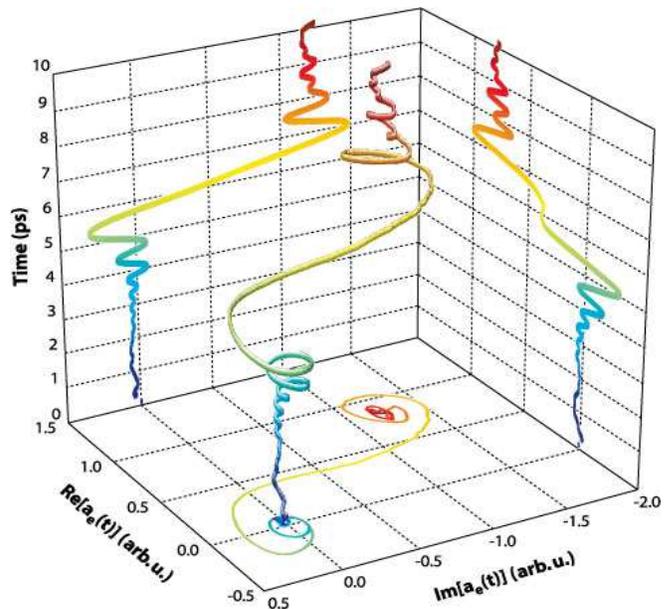 ,width=1\linewidth, height=8cm}
\caption[spirale3D] {Experimental probability amplitude
reconstructed from the CTs of Fig. \ref{Fig-CT}. Real part and
Imaginary parts are shown as a function of time.}
\label{Fig-spirale3D}
\end{center}
\end{figure}

The excited state probability amplitude produced by the second pulse
is extracted from the two measurements displayed in Fig.
\ref{Fig-CT}. The good stability of the laser allows the raw data to
be used without any renormalization between the two pairs of
measurements. The reconstructed probability amplitude is displayed
on Fig. \ref{Fig-spirale3D} in a 3D plot (complex plane as a
function of time). The projections on the various 2D planes are also
displayed. The expected Cornu spiral is clearly seen in the complex
plane.

\begin{figure}[!ht]
\begin{center}
\epsfig{figure=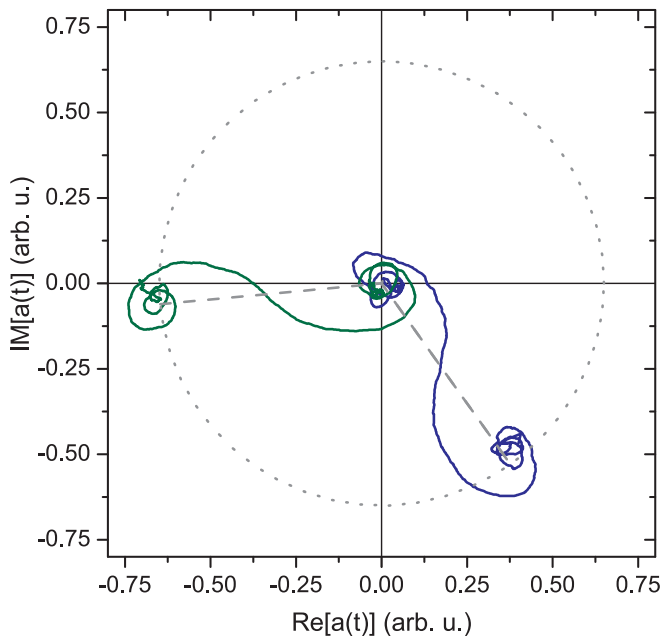,width=1\linewidth} \caption[compplane]
{Reconstructed experimental probability amplitude obtained for
different values of the relative phase : $\theta_0=3.3$ rad and
$\theta_0+2\pi/3$.} \label{Fig-compplane}
\end{center}
\end{figure}

The first part of the spiral winds around the population induced by
the first pulse (which has been substracted on the graph). After
passage through resonance corresponding to an almost "straight"
direction as expected (stationary phase in Eq.
\ref{cc:eq:apet-def}), the second part of the spiral winds around
the asymptotic value resulting from the combined effects of both
pulses.
%Because all the effects presented here are due to interference phenomena, they are extremely sensitive to the phase between the two pulses and one can describes all the complex plane by just varying the phase between the two pulses as shown on
The reconstruction procedure requires resolving a set of nonlinear
equations. Two solutions are in general mathematically obtained, but
only one is physically acceptable. It is obtained by continuity from
the solution corresponding to the beginning of the second pulse for
which $\aped(-\infty)=0$. This procedure works efficiently as long
as these two solutions are not degenerate. This may occur for
particular values of $\theta$ if the second pulse is larger than the
first one. This situation should therefore be avoided. However, a
scheme with three measurements as depicted above (see Fig.
\ref{FigCTtheo}) could instead be used in this situation. Figure
\ref{Fig-compplane} gives examples of Cornu spirals reconstructed
for $\theta$ values of $\theta_0=3.3$ rad and $\theta_0+2\pi/3$ for
similar pulse amplitudes. The quality of the reconstruction is
excellent in the two cases.

The present experiment benefits from the wide capability of our high
resolution pulse shaper \cite{pulseshaperRSI04}. In particular it
delivers a sequence of two shaped pulses with excellent
(interferometric) stability. The lack of stability was a serious
limitation (partly bypassed by measuring the noise induced by delay
fluctuations \cite{Averbukh_COIN95PRL,Bucksbaum_QSholography_PRL98})
in previous experiments \cite{Blanch97Cs}. This was overcome only
recently in Michelson type experiments \cite{Ohmori03,Katsuki06}
%,KatsukiPRL06}
.

The present scheme achieves time resolved measurement of a single
quantum state. Extension to a superposition of quantum states is
quite straightforward. The phase and amplitude pulse shaper allows
specific phases to be applied to the second pulse at the transition
frequencies associated with each independent excited state. A simple
scheme with $(2p+1)$ measurements for $p$ excited states can be
derived. One measurement can be used to remove the nonlinear term in
Eq. \ref{cc:eq:cttheta-final}, for instance with the second pulse
alone \cite{MonmayrantCT-spirograph-optCom06}. This provides a
system of $2p$ linear equations with $2p$ unknowns
%(real and imaginary part of each probability amplitude).
The set of specific
phases can be easily chosen to ensure that the corresponding system
can be inverted. Such a scheme would present an extension of quantum
state holography \cite{SchleichShapiro98PRL,SchleichShapiro99PRA} to
the time evolution holography of a complex quantum wave packet.

Weak-field interaction is not an intrinsic limitation of this
technique. Extension to the intermediate field regime is currently
under investigation. In this regime, iterative algorithms have been
proposed to reconstruct the quantum state in wave packet
interferences \cite{Yeazell_strongfield_reconstruction99} and could
be implemented in our scheme.

%The route has been shown in wave packet interferences on how to take into account
%A similar approach to In this case,

We have demonstrated how the instantaneous excited state can be
reconstructed during its interaction with a laser pulse. This scheme
is very general and can be used even without chirping the pulse. It
can have wide applications in situations where careful control of
the wave function during ultrashort pulses interaction is sought.
This is, in particular, the case in quantum information
applications. It can of course also be used to characterize complex
pulse shapes \cite{MonmayrantCT-reconstruction-05}.

%!!dans les autres ouvertures jepense qu'il faudraitun comment sur la phase absolue du champ est ce qu'on aurait une possibilité d'y entre sensible

We sincerely acknowledge Julien Courteaud for his help during a part
of the experiment and Chris Meier for fruitful discussions.
¨%\include{biblio}
\bibliographystyle{apsrev}
\bibliography{PRLspiralBib}

\begin{thebibliography}{26}
\expandafter\ifx\csname natexlab\endcsname\relax\def\natexlab#1{#1}\fi
\expandafter\ifx\csname bibnamefont\endcsname\relax
  \def\bibnamefont#1{#1}\fi
\expandafter\ifx\csname bibfnamefont\endcsname\relax
  \def\bibfnamefont#1{#1}\fi
\expandafter\ifx\csname citenamefont\endcsname\relax
  \def\citenamefont#1{#1}\fi
\expandafter\ifx\csname url\endcsname\relax
  \def\url#1{\texttt{#1}}\fi
\expandafter\ifx\csname urlprefix\endcsname\relax\def\urlprefix{URL }\fi
\providecommand{\bibinfo}[2]{#2}
\providecommand{\eprint}[2][]{\url{#2}}

\bibitem[{\citenamefont{Raymer}(1997)}]{Raymer97}
\bibinfo{author}{\bibfnamefont{M.~G.} \bibnamefont{Raymer}},
  \bibinfo{journal}{Contemp. Phys.} \textbf{\bibinfo{volume}{38}},
  \bibinfo{pages}{343} (\bibinfo{year}{1997}).

\bibitem[{\citenamefont{Schleich}(2001)}]{Schleichbook01}
\bibinfo{author}{\bibfnamefont{W.~P.} \bibnamefont{Schleich}},
  \emph{\bibinfo{title}{Quantum Optics in Phase Space}}
  (\bibinfo{publisher}{Wiley-VCH}, \bibinfo{address}{Berlin},
  \bibinfo{year}{2001}).

\bibitem[{\citenamefont{Leichtle et~al.}(1998)\citenamefont{Leichtle, Schleich,
  Averbukh, and Shapiro}}]{SchleichShapiro98PRL}
\bibinfo{author}{\bibfnamefont{C.}~\bibnamefont{Leichtle}},
  \bibinfo{author}{\bibfnamefont{W.~P.} \bibnamefont{Schleich}},
  \bibinfo{author}{\bibfnamefont{I.~S.} \bibnamefont{Averbukh}},
  \bibnamefont{and} \bibinfo{author}{\bibfnamefont{M.}~\bibnamefont{Shapiro}},
  \bibinfo{journal}{Phys. Rev. Lett.} \textbf{\bibinfo{volume}{80}},
  \bibinfo{pages}{1418} (\bibinfo{year}{1998}).

\bibitem[{\citenamefont{Averbukh et~al.}(1999)\citenamefont{Averbukh, Shapiro,
  Leichtle, and Schleich}}]{SchleichShapiro99PRA}
\bibinfo{author}{\bibfnamefont{I.~S.} \bibnamefont{Averbukh}},
  \bibinfo{author}{\bibfnamefont{M.}~\bibnamefont{Shapiro}},
  \bibinfo{author}{\bibfnamefont{C.}~\bibnamefont{Leichtle}}, \bibnamefont{and}
  \bibinfo{author}{\bibfnamefont{W.~P.} \bibnamefont{Schleich}},
  \bibinfo{journal}{Phys. Rev. A} \textbf{\bibinfo{volume}{59}},
  \bibinfo{pages}{2163} (\bibinfo{year}{1999}).

\bibitem[{\citenamefont{Chen and Yeazell}(1997)}]{Yeazell_holography_97}
\bibinfo{author}{\bibfnamefont{X.}~\bibnamefont{Chen}} \bibnamefont{and}
  \bibinfo{author}{\bibfnamefont{J.~A.} \bibnamefont{Yeazell}},
  \bibinfo{journal}{Phys. Rev. A} \textbf{\bibinfo{volume}{56}},
  \bibinfo{pages}{2316} (\bibinfo{year}{1997}).

\bibitem[{\citenamefont{Weinacht et~al.}(1998)\citenamefont{Weinacht, Ahn, and
  Bucksbaum}}]{Bucksbaum_QSholography_PRL98}
\bibinfo{author}{\bibfnamefont{T.~C.} \bibnamefont{Weinacht}},
  \bibinfo{author}{\bibfnamefont{J.}~\bibnamefont{Ahn}}, \bibnamefont{and}
  \bibinfo{author}{\bibfnamefont{P.~H.} \bibnamefont{Bucksbaum}},
  \bibinfo{journal}{Phys. Rev. Lett.} \textbf{\bibinfo{volume}{80}},
  \bibinfo{pages}{5508} (\bibinfo{year}{1998}).

\bibitem[{\citenamefont{Weinacht et~al.}(1999)\citenamefont{Weinacht, Ahn, and
  Bucksbaum}}]{Bucksbaum_holography_Nature99}
\bibinfo{author}{\bibfnamefont{T.~C.} \bibnamefont{Weinacht}},
  \bibinfo{author}{\bibfnamefont{J.}~\bibnamefont{Ahn}}, \bibnamefont{and}
  \bibinfo{author}{\bibfnamefont{P.~H.} \bibnamefont{Bucksbaum}},
  \bibinfo{journal}{Nature} \textbf{\bibinfo{volume}{397}},
  \bibinfo{pages}{233 } (\bibinfo{year}{1999}).

\bibitem[{\citenamefont{Dunn et~al.}(1995)\citenamefont{Dunn, Walmsley, and
  Mukamel}}]{Walmsley95}
\bibinfo{author}{\bibfnamefont{T.~J.} \bibnamefont{Dunn}},
  \bibinfo{author}{\bibfnamefont{I.~A.} \bibnamefont{Walmsley}},
  \bibnamefont{and} \bibinfo{author}{\bibfnamefont{S.}~\bibnamefont{Mukamel}},
  \bibinfo{journal}{Phys. Rev. Lett.} \textbf{\bibinfo{volume}{74}},
  \bibinfo{pages}{884} (\bibinfo{year}{1995}).

\bibitem[{\citenamefont{Leonhardt and Raymer}(1996)}]{Raymer96}
\bibinfo{author}{\bibfnamefont{U.}~\bibnamefont{Leonhardt}} \bibnamefont{and}
  \bibinfo{author}{\bibfnamefont{M.~G.} \bibnamefont{Raymer}},
  \bibinfo{journal}{Phys. Rev. Lett.} \textbf{\bibinfo{volume}{76}},
  \bibinfo{pages}{1985} (\bibinfo{year}{1996}).

\bibitem[{\citenamefont{Zucchetti et~al.}(1999)\citenamefont{Zucchetti, Vogel,
  Welsch, and Walmsley}}]{walmsleyPRA99heterodyne}
\bibinfo{author}{\bibfnamefont{A.}~\bibnamefont{Zucchetti}},
  \bibinfo{author}{\bibfnamefont{W.}~\bibnamefont{Vogel}},
  \bibinfo{author}{\bibfnamefont{D.~G.} \bibnamefont{Welsch}},
  \bibnamefont{and} \bibinfo{author}{\bibfnamefont{I.~A.}
  \bibnamefont{Walmsley}}, \bibinfo{journal}{Phys. Rev. A}
  \textbf{\bibinfo{volume}{60}}, \bibinfo{pages}{2716} (\bibinfo{year}{1999}).

\bibitem[{\citenamefont{Skovsen et~al.}(2003)\citenamefont{Skovsen,
  Stapelfeldt, Juhl, and Molmer}}]{Stapelfeldt03I2tomography}
\bibinfo{author}{\bibfnamefont{E.}~\bibnamefont{Skovsen}},
  \bibinfo{author}{\bibfnamefont{H.}~\bibnamefont{Stapelfeldt}},
  \bibinfo{author}{\bibfnamefont{S.}~\bibnamefont{Juhl}}, \bibnamefont{and}
  \bibinfo{author}{\bibfnamefont{K.}~\bibnamefont{Molmer}},
  \bibinfo{journal}{Phys. Rev. Lett.} \textbf{\bibinfo{volume}{91}},
  \bibinfo{pages}{090406} (\bibinfo{year}{2003}).

\bibitem[{\citenamefont{Broers et~al.}(1992)\citenamefont{Broers, van Linden
  van~den Heuvell, and Noordam}}]{Noordamladder92}
\bibinfo{author}{\bibfnamefont{B.}~\bibnamefont{Broers}},
  \bibinfo{author}{\bibfnamefont{H.~B.} \bibnamefont{van Linden van~den
  Heuvell}}, \bibnamefont{and} \bibinfo{author}{\bibfnamefont{L.~D.}
  \bibnamefont{Noordam}}, \bibinfo{journal}{Phys. Rev. Lett.}
  \textbf{\bibinfo{volume}{69}}, \bibinfo{pages}{2062} (\bibinfo{year}{1992}).

\bibitem[{\citenamefont{Vitanov}(1999)}]{Vitanov99}
\bibinfo{author}{\bibfnamefont{N.~V.} \bibnamefont{Vitanov}},
  \bibinfo{journal}{Phys. Rev. A} \textbf{\bibinfo{volume}{59}},
  \bibinfo{pages}{988} (\bibinfo{year}{1999}).

\bibitem[{\citenamefont{Zamith et~al.}(2001)\citenamefont{Zamith, Degert,
  Stock, de~Beauvoir, Blanchet, Bouchene, and Girard}}]{zamith01}
\bibinfo{author}{\bibfnamefont{S.}~\bibnamefont{Zamith}},
  \bibinfo{author}{\bibfnamefont{J.}~\bibnamefont{Degert}},
  \bibinfo{author}{\bibfnamefont{S.}~\bibnamefont{Stock}},
  \bibinfo{author}{\bibfnamefont{B.}~\bibnamefont{de~Beauvoir}},
  \bibinfo{author}{\bibfnamefont{V.}~\bibnamefont{Blanchet}},
  \bibinfo{author}{\bibfnamefont{M.~A.} \bibnamefont{Bouchene}},
  \bibnamefont{and} \bibinfo{author}{\bibfnamefont{B.}~\bibnamefont{Girard}},
  \bibinfo{journal}{Phys. Rev. Lett.} \textbf{\bibinfo{volume}{87}},
  \bibinfo{pages}{033001} (\bibinfo{year}{2001}).

\bibitem[{\citenamefont{Rothenberg}(1986)}]{Rothenberg86}
\bibinfo{author}{\bibfnamefont{J.~E.} \bibnamefont{Rothenberg}},
  \bibinfo{journal}{IEEE J. Quant. Electronics}
  \textbf{\bibinfo{volume}{QE-22}}, \bibinfo{pages}{174}
  (\bibinfo{year}{1986}).

\bibitem[{\citenamefont{Delagnes et~al.}(2005)\citenamefont{Delagnes,
  Monmayrant, Zahariev, Chatel, Girard, and Bouchene}}]{Delagnes04}
\bibinfo{author}{\bibfnamefont{J.}~\bibnamefont{Delagnes}},
  \bibinfo{author}{\bibfnamefont{A.}~\bibnamefont{Monmayrant}},
  \bibinfo{author}{\bibfnamefont{P.}~\bibnamefont{Zahariev}},
  \bibinfo{author}{\bibfnamefont{B.}~\bibnamefont{Chatel}},
  \bibinfo{author}{\bibfnamefont{B.}~\bibnamefont{Girard}}, \bibnamefont{and}
  \bibinfo{author}{\bibfnamefont{M.~A.} \bibnamefont{Bouchene}},
  \bibinfo{journal}{Appl. Phys. B} \textbf{\bibinfo{volume}{submitted}}
  (\bibinfo{year}{2005}).

\bibitem[{\citenamefont{Degert et~al.}(2002)\citenamefont{Degert, Wohlleben,
  Chatel, Motzkus, and Girard}}]{Degert02CTshaped}
\bibinfo{author}{\bibfnamefont{J.}~\bibnamefont{Degert}},
  \bibinfo{author}{\bibfnamefont{W.}~\bibnamefont{Wohlleben}},
  \bibinfo{author}{\bibfnamefont{B.}~\bibnamefont{Chatel}},
  \bibinfo{author}{\bibfnamefont{M.}~\bibnamefont{Motzkus}}, \bibnamefont{and}
  \bibinfo{author}{\bibfnamefont{B.}~\bibnamefont{Girard}},
  \bibinfo{journal}{Phys. Rev. Lett.} \textbf{\bibinfo{volume}{89}},
  \bibinfo{pages}{203003} (\bibinfo{year}{2002}).

\bibitem[{\citenamefont{Wohlleben et~al.}(2004)\citenamefont{Wohlleben, Degert,
  Monmayrant, Chatel, Motzkus, and Girard}}]{RbShapingAPB04}
\bibinfo{author}{\bibfnamefont{W.}~\bibnamefont{Wohlleben}},
  \bibinfo{author}{\bibfnamefont{J.}~\bibnamefont{Degert}},
  \bibinfo{author}{\bibfnamefont{A.}~\bibnamefont{Monmayrant}},
  \bibinfo{author}{\bibfnamefont{B.}~\bibnamefont{Chatel}},
  \bibinfo{author}{\bibfnamefont{M.}~\bibnamefont{Motzkus}}, \bibnamefont{and}
  \bibinfo{author}{\bibfnamefont{B.}~\bibnamefont{Girard}},
  \bibinfo{journal}{App. Phys. B} \textbf{\bibinfo{volume}{79}},
  \bibinfo{pages}{435} (\bibinfo{year}{2004}).

\bibitem[{\citenamefont{Monmayrant et~al.}(2006)\citenamefont{Monmayrant,
  Chatel, and Girard}}]{MonmayrantCT-reconstruction-05}
\bibinfo{author}{\bibfnamefont{A.}~\bibnamefont{Monmayrant}},
  \bibinfo{author}{\bibfnamefont{B.}~\bibnamefont{Chatel}}, \bibnamefont{and}
  \bibinfo{author}{\bibfnamefont{B.}~\bibnamefont{Girard}},
  \bibinfo{journal}{Opt. Lett.} \textbf{\bibinfo{volume}{31}},
  \bibinfo{pages}{xxx} (\bibinfo{year}{2006}).

\bibitem[{\citenamefont{Monmayrant and Chatel}(2004)}]{pulseshaperRSI04}
\bibinfo{author}{\bibfnamefont{A.}~\bibnamefont{Monmayrant}} \bibnamefont{and}
  \bibinfo{author}{\bibfnamefont{B.}~\bibnamefont{Chatel}},
  \bibinfo{journal}{Rev. Sci. Inst.} \textbf{\bibinfo{volume}{75}},
  \bibinfo{pages}{2668} (\bibinfo{year}{2004}).

\bibitem[{\citenamefont{Kinrot et~al.}(1995)\citenamefont{Kinrot, Averbukh, and
  Prior}}]{Averbukh_COIN95PRL}
\bibinfo{author}{\bibfnamefont{O.}~\bibnamefont{Kinrot}},
  \bibinfo{author}{\bibfnamefont{I.}~\bibnamefont{Averbukh}}, \bibnamefont{and}
  \bibinfo{author}{\bibfnamefont{Y.}~\bibnamefont{Prior}},
  \bibinfo{journal}{Phys. Rev. Lett.} \textbf{\bibinfo{volume}{75}},
  \bibinfo{pages}{3822} (\bibinfo{year}{1995}).

\bibitem[{\citenamefont{Blanchet et~al.}(1997)\citenamefont{Blanchet, Nicole,
  Bouchene, and Girard}}]{Blanch97Cs}
\bibinfo{author}{\bibfnamefont{V.}~\bibnamefont{Blanchet}},
  \bibinfo{author}{\bibfnamefont{C.}~\bibnamefont{Nicole}},
  \bibinfo{author}{\bibfnamefont{M.~A.} \bibnamefont{Bouchene}},
  \bibnamefont{and} \bibinfo{author}{\bibfnamefont{B.}~\bibnamefont{Girard}},
  \bibinfo{journal}{Phys. Rev. Lett.} \textbf{\bibinfo{volume}{78}},
  \bibinfo{pages}{2716} (\bibinfo{year}{1997}).

\bibitem[{\citenamefont{Ohmori et~al.}(2003)\citenamefont{Ohmori, Sato,
  Nikitin, and Rice}}]{Ohmori03}
\bibinfo{author}{\bibfnamefont{K.}~\bibnamefont{Ohmori}},
  \bibinfo{author}{\bibfnamefont{Y.}~\bibnamefont{Sato}},
  \bibinfo{author}{\bibfnamefont{E.~E.} \bibnamefont{Nikitin}},
  \bibnamefont{and} \bibinfo{author}{\bibfnamefont{S.~A.} \bibnamefont{Rice}},
  \bibinfo{journal}{Phys. Rev. Lett.} \textbf{\bibinfo{volume}{91}},
  \bibinfo{pages}{243003} (\bibinfo{year}{2003}).

\bibitem[{\citenamefont{Katsuki et~al.}(2006)\citenamefont{Katsuki, Chiba,
  Girard, Meier, and Ohmori}}]{Katsuki06}
\bibinfo{author}{\bibfnamefont{H.}~\bibnamefont{Katsuki}},
  \bibinfo{author}{\bibfnamefont{H.}~\bibnamefont{Chiba}},
  \bibinfo{author}{\bibfnamefont{B.}~\bibnamefont{Girard}},
  \bibinfo{author}{\bibfnamefont{C.}~\bibnamefont{Meier}}, \bibnamefont{and}
  \bibinfo{author}{\bibfnamefont{K.}~\bibnamefont{Ohmori}},
  \bibinfo{journal}{Science} \textbf{\bibinfo{volume}{311}},
  \bibinfo{pages}{xxx} (\bibinfo{year}{2006}).

\bibitem[{\citenamefont{Monmayrant et~al.}(2006S)\citenamefont{Monmayrant,
  Chatel, and Girard}}]{MonmayrantCT-spirograph-optCom06}
\bibinfo{author}{\bibfnamefont{A.}~\bibnamefont{Monmayrant}},
  \bibinfo{author}{\bibfnamefont{B.}~\bibnamefont{Chatel}}, \bibnamefont{and}
  \bibinfo{author}{\bibfnamefont{B.}~\bibnamefont{Girard}},
  \bibinfo{journal}{Opt. Commun.} \textbf{\bibinfo{volume}{submitted}}
  (\bibinfo{year}{2006S}).

\bibitem[{\citenamefont{Chen and
  Yeazell}(1999)}]{Yeazell_strongfield_reconstruction99}
\bibinfo{author}{\bibfnamefont{X.}~\bibnamefont{Chen}} \bibnamefont{and}
  \bibinfo{author}{\bibfnamefont{J.~A.} \bibnamefont{Yeazell}},
  \bibinfo{journal}{Phys. Rev. A} \textbf{\bibinfo{volume}{60}},
  \bibinfo{pages}{4253} (\bibinfo{year}{1999}).

\end{thebibliography}
%\begin{thebibliography}{99}
%\end{thebibliography}
\end{document}